\documentclass{aa}
\usepackage{txfonts}
\usepackage{graphicx}
\usepackage[normalem]{ulem}

\begin{document}

\title{Kinematics of young star clusters in the outer north-eastern region of the
Small Magellanic Cloud}

\author{Andr\'es E. Piatti\inst{1,2,}\thanks{\email{andres.piatti@fcen.uncu.edu.ar}}}

\institute{Instituto Interdisciplinario de Ciencias B\'asicas (ICB), CONICET-UNCuyo, Padre J. Contreras 1300, M5502JMA, Mendoza, Argentina;
\and Consejo Nacional de Investigaciones Cient\'{\i}ficas y T\'ecnicas (CONICET), Godoy Cruz 2290, C1425FQB,  Buenos Aires, Argentina\\
}

\date{Received / Accepted}

\abstract{It has been suggested since recent time that the magnitude of the
interaction between galaxies could be measured from the level of kinematic disturbance
of their outer regions with respect to the innermost ones. Here, I proved that the
outer north-eastern region of the Small Magellanic Cloud (SMC), a relatively recent
stellar structure with a tidal origin from the interaction with the Large Magellanic Cloud,
is imprinted by a residual velocity pattern. I obtained from GEMINI GMOS spectra 
mean radial velocities of star clusters formed in situ, which added to derived 
mean proper motions and heliocentric distances, allowed to compute their 3D
space velocity components. These space velocities differentiate from those
that the clusters would have if they instead orderly rotated with the galaxy, i.e.,
their residual velocities are larger than the upper limit for an object
pertaining to the SMC main body rotation disk. The level of kinematic disturbance
depends on the SMC rotation disk adopted; galaxy rotation disks traced using
relatively old objects are discouraged.The resulting kinematic disturbance
arises in younger and older stellar populations, so that the epoch of close
interaction between both Magellanic Clouds cannot be uncovered on the basis
of the kinematics behavior of stellar populations populating the outer SMC
regions.}

\keywords{techniques: spectroscopic -- galaxies: individual: SMC -- galaxies: star clusters}

\titlerunning{Kinematics of SMC star clusters}  

\authorrunning{A.E. Piatti}           

\maketitle

\markboth{A.E. Piatti:}{Kinematics of SMC star clusters}

\section{Introduction}           

In recent years, the tidal interaction between galaxies has become an issue of interest to 
understand their formation and evolution histories. One of the witnesses of such tidal 
interactions is the presence of an internal galactic kinematics pattern, which show increasing 
velocities dispersion towards the outermost galactic regions with respect to that in the 
innermost ones. For instance, \citet{martinezgarciaetal2023} found that dwarf spheroidal 
satellites of the Milky Way present velocity gradients along the line-of-sight, indicating that 
the interaction with the Milky Way is causing them. To draw such a conclusion they combined 
proper motions from the {\it Gaia} Data Release 3 (DR3) \citep{gaiaetal2016,gaiaetal2022b} 
and line-of-sight velocities from the literature to derive their 3D internal kinematics.

In this context, the outermost regions of the Small Magellanic Cloud (SMC), particularly those
facing towards the Large Magellanic Cloud (LMC), are expected to have experienced the effects 
of the mutual tidal interaction more intensively than its  stellar innermost component, 
which exhibit an orderly rotational motion \citep[see, e.g.][and references therein]{niederhoferetal2018,zivicketal2018,diteodoroetal2019}: the observed 
gaseous component not showing a evidence of ordered rotation 
\citep{murrayetal2019,rathoreetal2025} though. However, the velocity gradient for the 
SMC has not been estimated yet  (with the exception of one estimate by \citet{zivicketal2021}),
which is fundamental to contribute to our knowledge about the 
formation, evolution and interaction of these pair of galaxies.

\citet{martinezdelgadoetal2019} studied the nature of a stellar structure (also called shell region)
located
8$\degr$ north-east from the SMC, and concluded that it formed 
in a recent star formation 
event, likely triggered by the interaction with the LMC. They also showed that the only 9 star 
clusters projected onto that structure are young, compatible with being part of it. 
\citet{piatti2022d} used the Survey of the MAgellanic Stellar History 
\citep[SMASH;][]{nideveretal2021} data base to derive ages and distances of these star clusters, 
and confirmed that they are young and span a heliocentric distance range $\sim$ 3 times the depth 
of the SMC main body. Both, their youth and their large distance range along the same 
line-of-sight favor the tidal origin of this north-eastern stellar structure, making it suitable 
to investigate whether their kinematics differentiates from that of the rotating inner SMC main 
body.

If the kinematics of this outermost SMC region were affected by tidal effects, it should show 
a kinematics that departs from that of the rotating inner disk extrapolated toward this 
outermost region. Based on this, 3D velocities of star clusters aligned along the line-of-sight 
of the north-eastern stellar structure are important: 1) they add a valuable puzzle piece on 
the kinematics of the SMC at large distance from its center; 2) they help us to understand at 
what level the interaction of the SMC with the LMC affects its kinematics, and  3) they contribute 
to build a comprehensive map of the kinematic behavior of the 
SMC outermost regions, which is still a matter of work 
\citep[see, e.g., Sloan Digital Sky Survey V:][]{kollmeieretal2019}. Note that by kinematically 
studying these 
young clusters we improve our knowledge of the recent SMC formation, evolution and tidal 
interaction. Precisely, in this work I make use of spectroscopic and astrometric data to study 
the kinematics of star clusters in order to address this issue. Section~2 describes the collection 
and processing of the acquired spectroscopic data, Section~3 deals with the estimation of mean star
clusters' radial velocities, while Section~4 is devoted to the analysis of 
the derived 3D star cluster velocities. Section~5 summarizes the main conclusions of this work.

\section{Observational data}

I used the GEMINI GMOS-S spectrograph, in multi-object mode. to get spectra with the grism 
B1200 (central $\lambda$ $\sim$5000 \AA, spectral range 4200 - 5800 \AA, dispersion 0.26 \AA/pix) 
of stars in the fields of the SMC young star clusters HW~64, IC~1655, and IC~1660 
\citep[30-105 Myr;][]{piatti2022d}, respectively. Stars were selected based on 
the membership probabilities of \citet{piatti2022d}, including stars only brighter than $g$ = 20 mag. 
I obtained
3$\times$510 sec exposures using a manufactured mask for each star cluster field through program
GS-2025B-Q-101 (PI: Illesca). The collection of images included Cu-Ar arc lamps and 
flat-fields obtained before or after each individual science spectrum, as well as, nightly series 
of bias and observations of spectrophotometric standard stars. Figures~\ref{fig1} to \ref{fig3} 
show the positions in the sky of the observed stars and their loci in the SMASH color-magnitude 
diagram (CMD).

The spectra images were reduced using the GEMINI task package\footnote{https://www.gemini.edu/observing/phase-iii/reducing-data/gemini-iraf-data-reduction-software} within the 
IRAF\footnote{https://iraf-community.github.io/} environment. Science images were trimmed, 
corrected by overscan, bias, flat-field, and mosaiced. Then, the slits were properly cut so that 
the star spectra appeared centered on them. The mosaiced spectra images were wavelength calibrated,
from which the individual star spectra were extracted, previously being background subtracted, and 
finally they were flux calibrated. I combined the three individual spectra per star in order to
remove cosmic rays and increase the spectra S/N ratio. I employed the resulting spectra to measure
radial velocity (RV) using the H$\beta$ spectral line and the FXCOR task within IRAF. I applied
the corresponding correction to attain heliocentric RVs. I derived RVs for most of the observed
stars, whose values are listed in Table~\ref{tab1a}, alongside the respective S/N ratio, the
celestial equatorial coordinates and the  observed SMASH $g,i$ photometry.

\section{Mean star clusters' radial velocity}

In order to estimate the mean star clusters' RV, I took advantage of the fact that a cluster 
member not only has a RV similar to the mean value of the cluster, but also it is placed
along the star cluster sequence in the CMD \citep[see Figure~2 in ][]{piatti2022d}.
These two parameters, the RV and the distance to the ridge line of the cluster sequence in the
CMD, are characterized by showing overdensities of cluster members in that 2D plane. 
This means that any Gaussian mixture model can be applied to disentangle
cluster members from field stars. 

\begin{figure*}
\includegraphics[width=\columnwidth]{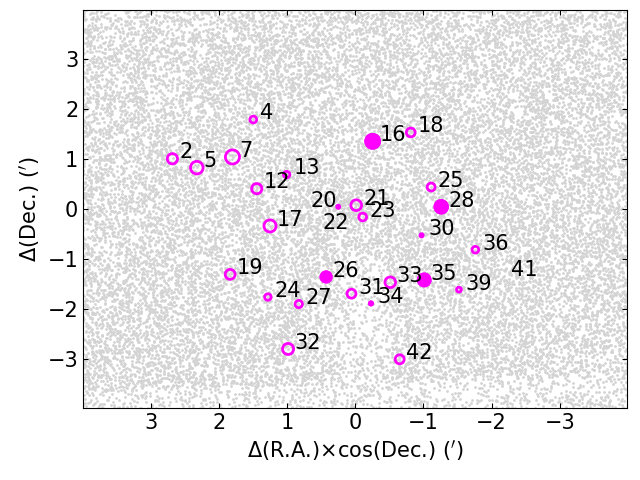}
\includegraphics[width=\columnwidth]{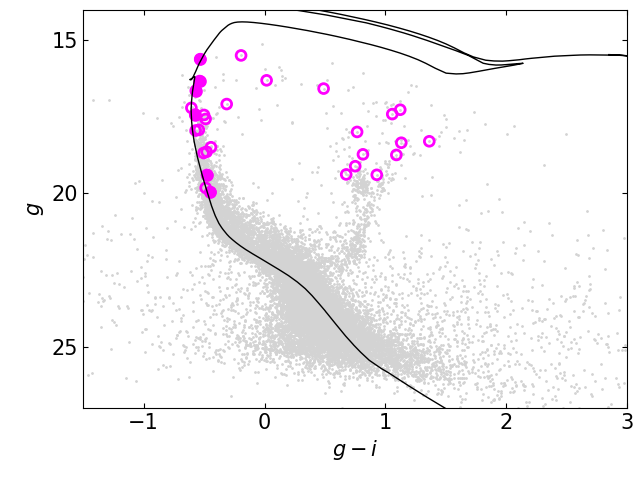}
\caption{{\it Left:} Sky map with stars observed by SMASH (gray points) and those
observed in this work (large magenta circles) in the field of HW~64; symbol size being 
proportional to the $g$ brightness of the star.  Filled magenta circles represent
cluster members (see details in Section~3). {\it Right:} SMASH color-magnitude
diagram with the isochrone \citep[][PARSEC v1.2S]{betal12} corresponding to the distance, 
reddening, metallicity and age of the cluster \citep{piatti2022d}, superimposed.} 
\label{fig1}
\end{figure*}

\begin{figure*}
\includegraphics[width=\columnwidth]{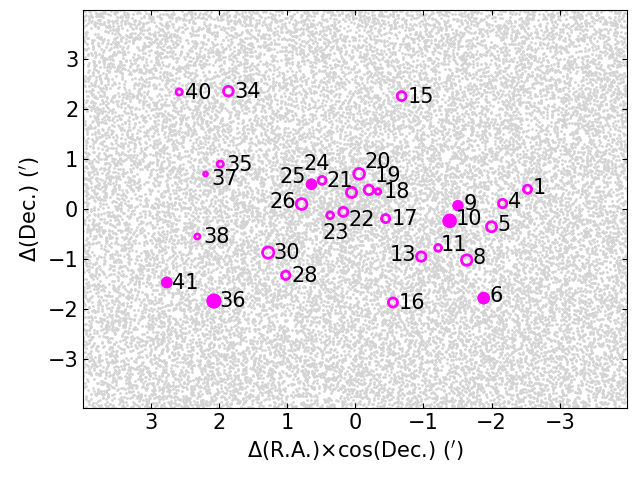}
\includegraphics[width=\columnwidth]{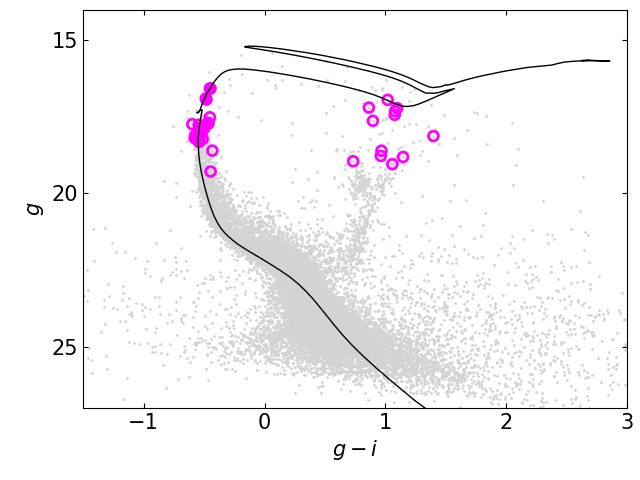}
\caption{Same as Figure~\ref{fig1} for IC~1655.}
\label{fig2}
\end{figure*}

\begin{figure*}
\includegraphics[width=\columnwidth]{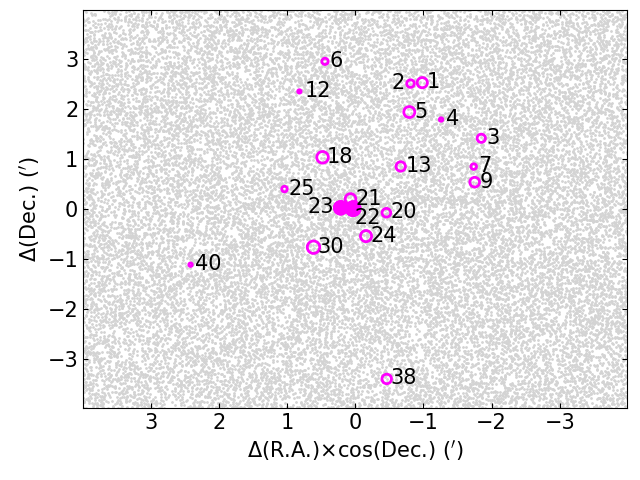}
\includegraphics[width=\columnwidth]{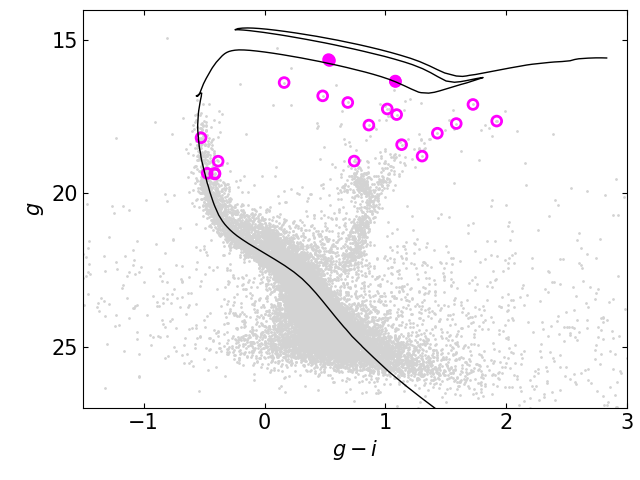}
\caption{Same as Figure~\ref{fig1} for IC~1660.}
\label{fig3}
\end{figure*}

I computed the distance of each observed star to the theoretical isochrone of the respective
cluster (see Figures~\ref{fig1} to \ref{fig3}) using the following expression:\\

$d$ = $\sqrt{((g-i) - (g-i)_{\rm iso})^2 + k\times(g - g_{\rm iso})^2}$\\

\noindent where $g$ and $(g-i)$ are the SMASH observed magnitudes and colors (see Table~\ref{tab1a})
and $g_{\rm iso}$ and $(g-i)_{\rm iso}$ are the theoretical ones with the respective cluster reddening 
and distance modulus applied. I employed the cluster parameters (age, reddening, distance and metallicity) 
derived by \citet{piatti2022d}. The parameter $k$ is a factor that put magnitudes and colors in the same 
scale. In this case, I adopted $k$= 0.5. In practice, for each ($g$, $g-i$) pair I computed $d$ using many 
points along the theoretical isochrone, and chose the $g_{\rm iso}$ and  $(g-i)_{\rm iso}$ values that 
minimized $d$. For the sake of the reader, I called $d$ the scaled CMD distance. The uncertainties in 
$d$ were computed by propagating the errors in $g$ and $g-i$.

\begin{figure*}
\includegraphics[width=\textwidth/3]{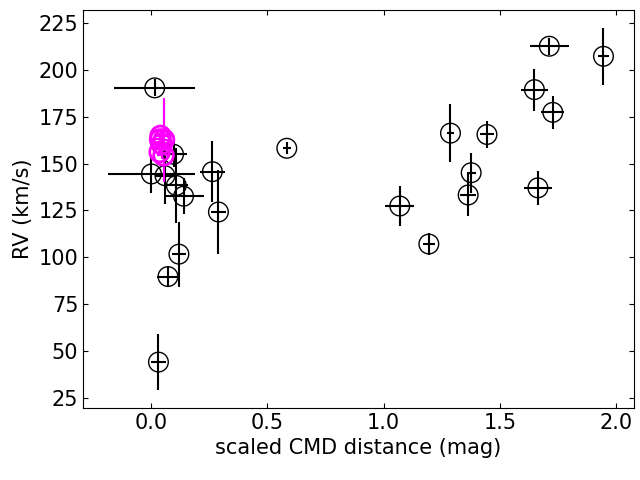}
\includegraphics[width=\textwidth/3]{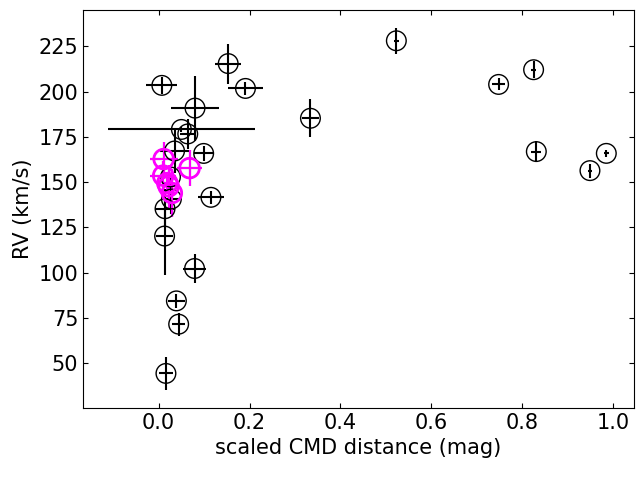}
\includegraphics[width=\textwidth/3]{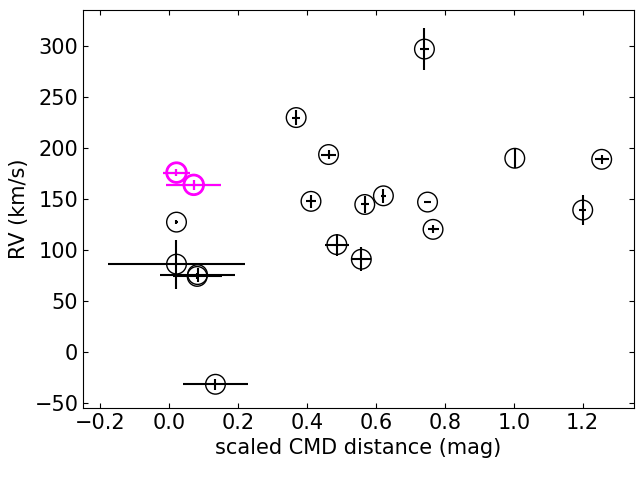}
\caption{RV versus scaled CMD distance of stars in HW~64 (left panel),
IC~1655 (middle panel), and IC~1660 (right panel),
respectively. The magenta circles represent the selected clustered stars.}
\label{fig4}
\end{figure*}

Figure~\ref{fig4} depicts the distribution of the measured stars in the RV versus scaled CMD distance
plane. As can be seen, the observed sample of stars includes field stars spanning the whole ranges
of RV and scaled CMD distance, and groups of stars that resemble those of cluster members. Field
stars are seen placed along the cluster line-of-sight, mingled with cluster members, but
having RVs different from that of cluster's stars and positions in the CMD out of the cluster
sequences. Likewise, it is possible to find field stars with RVs similar to that of the cluster or
close to the cluster's sequences in the CMD.

In order to detect any overdensity in the panels of Figure~\ref{fig4}, which in turn provide the
mean clusters' RVs, I used the Hierarchical Density-Based Spatial Clustering of Applications with 
Noise \citep[HDBSCAN][]{mcinnesetal2017} Gaussian mixture model technique.
The \texttt{min$\_$cluster$\_$size} parameter was varied between 3 and 8 dex in steps of 1 dex,
until reached an optimum value where the number of selected stars remained constant. Since 
each observed cluster field contains only one star cluster, the respective RV versus scaled CMD distance 
plane should consequently harbor only one overdensity. For this reason, I considered 
\texttt{allow$\_$single$\_$cluster=True} during the execution of HDBSCAN. Figure~\ref{fig4} shows 
with magenta circles the stars found by HDBSCAN that exhibit clustering.  I also highlight them from all 
the measures stars in Figures~\ref{fig1} to \ref{fig3} with filled magenta symbols. As can be seen, 
all of them are within 0.10 mag from the corresponding CMD cluster sequence. From the selected 
clustered stars I estimated the mean clusters' RVs.  Table~\ref{tab1} lists the reddening ($E(B-V)$),
true distance modulus ($m-M_o$), age, and metallicity (Z) values 
for the three studied clusters, taken from \citet{piatti2022d}, together with the mean cluster
RVs and the final number of confirmed cluster members identified from which the mean RVs 
are calculated. As far as I am aware, there are no previous
mean star cluster RV estimates.

\begin{table*}
\centering
\caption{Astrophysical properties of studied SMC star clusters.}
\label{tab1}
\begin{tabular}{lccccccccc}
\hline\hline
Name & $E(B-V)$ & $m-M_o$ & log(age /yr) & Z & RV & N$_{RV}$ & pmra & pmdec & N$_{pm}$ \\
          &  (mag)     &  (mag)      &                    & (dex)  & (km s$^{-1}$) & & (mas/yr) &  (mas/yr) & \\\hline
HW~64 & 0.142$\pm$0.026 & 19.193$\pm$0.083 & 7.468$\pm$0.309 & 0.0064$\pm$0.0029 & 160.24$\pm$3.75 & 6 & 0.812$\pm$0.032 & -1.232$\pm$0.028 & 6 \\
IC~1655 & 0.129$\pm$0.012 & 18.659$\pm$0.039 & 8.054$\pm$0.107 & 0.0037$\pm$0.0015 & 152.52$\pm$6.29 & 6 & 0.857$\pm$0.098 & -1.215$\pm$0.065 & 6 \\
IC~1660 & 0.139$\pm$0.017 & 18.813$\pm$0.055 & 7.909$\pm$0.143 & 0.0054$\pm$0.0023 & 170.20$\pm$6.00 & 2 & 0.965$\pm$0.112 & -1.242$\pm$0.054 & 6 \\\hline
\end{tabular}
\end{table*}

\section{Analysis and discussion}

With the aim of computing the 3D velocity components of HW~64. IC~1655, and IC~1660, I first derived the
mean clusters' proper motions. For that purpose, I download from {\it Gaia} DR3 R.A. and Dec. 
coordinates, parallaxes ($\varpi$), proper motions in R.A. and Dec. (pmra, pmdec), with their associated
uncertainties, and $G$, $BP$, and $RP$ magnitudes of stars located inside circles with a radius of 4 arcmin 
from the centers of these three star clusters. I restricted the data to those with errors in both
proper motions $<$ 0.1 mas/yr, \texttt{ruwe} $<$ 1.4 \citep{ripepietal2019},  and excluded confirmed 
nearby stars, i.e., $|\varpi|$ $<$ 3$\sigma(\varpi)$ \citep[see][]{vasiliev2018}. Then, I built
the clusters' CMDs and measured the scaled CMD distances $d$ as in Section~3, using the {\it Gaia}
theoretical isochrones \citep[][PARSEC v1.2S]{betal12}. Finally, I fed HDBSCAN with the collected data for
three variables: pmra, pmdec, and $d$, and identified the groups of stars satisfying cluster
overdensity in that 3D phase space. The resulting mean clusters' proper motions
are listed in Table~\ref{tab1}, alongside with the number of stars used to compute them.

\citet{piatti2021b} used derived proper motions and radial velocities
available in the literature of 25 SMC star clusters and the transformation 
Eqs. (9), (13), and (21) in \citet{vdmareletal2002} to search for the kinematic model 
that best represents the motion of that cluster sample. He fit the
R.A. and Dec., the heliocentric distance, the proper motion and the
systemic velocity of the SMC center, in addition to the inclination, the
position angle of the line-of-nodes, and the rotation velocity of the
SMC disk. Table~\ref{tab2} lists the derived values. I here employed the
transformation equations and the rotation model mentioned above to compute both, 
the 3D space velocity components of HW~64, IC~1655, and IC~1660, and those
corresponding to points on  the nominal SMC disk located at the clusters' distances
from the SMC center. In doing that, I used the obtained mean clusters' RVs and
proper motions and their heliocentric distances ($D$) estimated by \citet{piatti2022d}.
The uncertainties were calculated using the measured errors in proper motion and radial 
velocity, and those from the model solution for the 3D movement of the SMC center, 
propagated through the transformation equations and added in quadrature.

\begin{figure*}
\includegraphics[width=\textwidth/3]{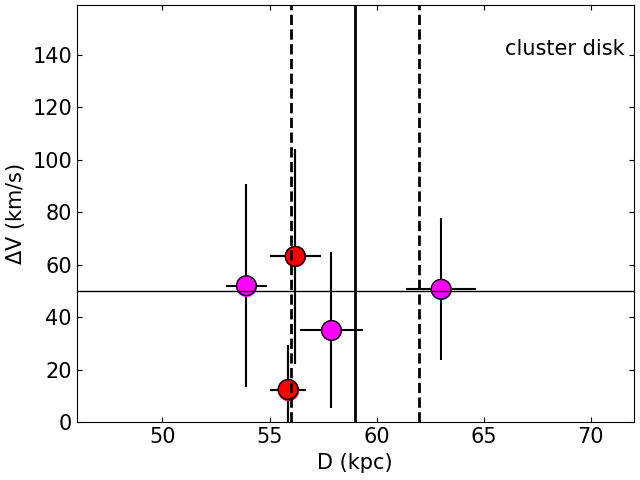}
\includegraphics[width=\textwidth/3]{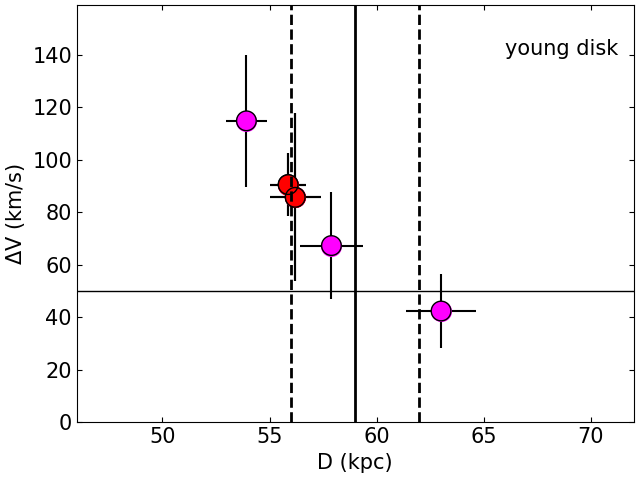}
\includegraphics[width=\textwidth/3]{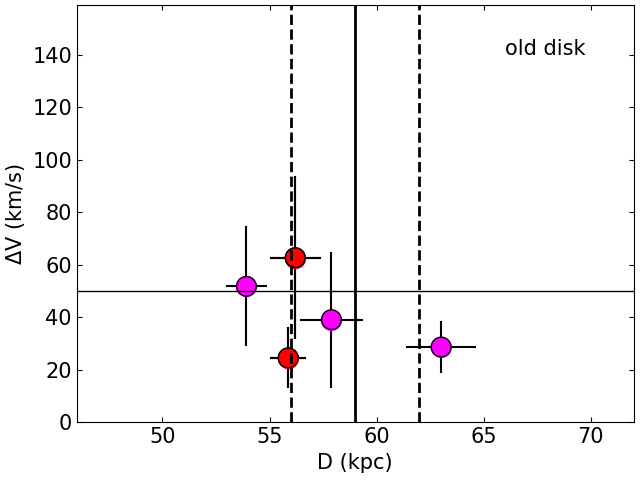}
\caption{$\Delta$$V$ versus heliocentric distance of young SMC clusters
located in the north-eastern shell region. The solid and
dashed vertical lines represent the galaxy center and its depth according to the
cluster rotation disk \citep{piatti2021b}, while the horizontal line represents
a boundary between bound and kinematically perturbed clusters.
Magenta and red filled circles represent the present studied clusters and
those taken from the literature, respectively.
A kinematic gradient is present when compared to a young stellar model
(middle panel), but 
absent in comparison to the old stellar and cluster models (right and left panels), respectively
 (see text for details).}
\label{fig5}
\end{figure*}

When both velocity vectors are subtracted one with
respect to the other and calculated the modulus of the resulting vector, an
index called residual velocity ($\Delta$$V$) is obtained \citep{piatti2021f}. 
$\Delta$$V$ is meant to measure the magnitude of kinematic disturbance of
star clusters with respect to an orderly rotation galaxy disk. 
Figure~\ref{fig5}, left panel, shows $\Delta$$V$ versus $D$ for the three
studied clusters represented with filled magenta circles ($D$= 63.0, 53.9, 57.9 kpc
for HW~64, IC~1655, and IC~1660, respectively). I enlarged the sample of
clusters genuinely belonging to the outer north-eastern region of the SMC
by searching the literature for the necessary information to compute
$\Delta$$V$. Thus, I added HW~56  \citep[3.09 Gyr][]{diasetal2021} and NGC~458 \citep[0.14 Gyr][]{songetal2021},
which are painted with filled red circles in Figure~\ref{fig5}. 
Bearing in mind the SMC center and line-of-sight depth obtained by \citet{piatti2021b},
which are represented by a solid and dashed vertical lines in Figure~\ref{fig5}, 
respectively, and the boundary between bound and kinematically perturbed clusters 
\citep[$\Delta$$V$ $\sim$ 60 km s$^{-1}$;][]{piatti2021f}, the five star clusters pertaining 
to the outer north-eastern SMC structure would not seem to be perturbed by tidal effects
from the LMC. Their $\Delta$$V$ values are on average below $\sim$ 60 km s$^{-1}$.

However, $\Delta$$V$ depends on the SMC rotation disk adopted, so that a comprehensive
analysis of the kinematic of the studied star clusters requires the consideration of
different rotation disk models. Recently, \citet[][and references therein]{dhanushetal2025} 
used {\it Gaia} DR3 data sets to derive kinematic parameters of different SMC stellar 
populations. Indeed, from young to old stellar population they found a change in the
SMC disk inclination from $\sim$82$\degr$ down to $\sim$58$\degr$, and in the position
angle of the line-of-nodes (LON) from $\sim$180$\degr$ up to $\sim$240$\degr$. 
 Table~\ref{tab2} includes the values of the parameters adopted in this work for a 
representative young and old stellar population, making use of \citet{dhanushetal2025}'s
results. As for the SMC center proper motion, I adopted that of \citet{niederhoferetal2021}
for the optical center \citep{dVF1972}, which is in very good agreement with those from recent 
space-based data \citep{gaiaetal2018b,zivicketal2018,deleoetal2020} and with \citet{dhanushetal2025}.

For these two representative SMC rotation disks, I computed $\Delta$$V$ and their uncertainties
for the five studied star clusters in the outer north-eastern structure, and the resulting
$\Delta$$V$ versus $D$ relationships are depicted in Figure~\ref{fig5} (middle and right panels).
As can be seen, the kinematics of the studied clusters with respect to the orderly motion of the
young stellar population \citep{niederhoferetal2018,zivicketal2018,diteodoroetal2019} would
seem to reveal a kinematic gradient, in the sense that smaller the heliocentric distance
(which means the closer to the LMC), the larger the $\Delta$$V$ value. This outcome
suggests that tidal interactions with the LMC (the studied clusters are relatively young)
could have imprinted certain agitation in their kinematics. This possibility would seem to
be supported by the fact that the correlation of $\Delta$$V$ and $D$ using the old star
population kinematic model resulted to be similar to that from the cluster rotation model.
Indeed, the 25 SMC clusters used by \citet{piatti2021b} to uncover their own kinematic
model are intermediate-age to old globular clusters (age $>$ 1 Gyr). In other words,
the old star population rotation disk is based on the motions of perturbed
stars located in the outer SMC regions, and hence no readily visible gradient is seen in
the left and right panels of Figure~\ref{fig5}.

\begin{table*}
\centering
\caption{SMC rotation disk models.}
\label{tab2}
\begin{tabular}{lccc}
\hline\hline
Parameter                           &    cluster disk  & young disk (age $<$ 50 Myr) & old disk (age > 2 Gyr) \\\hline
SMC center R.A. ($\degr$):          &   13.30$\pm$0.10  & 13.05 & 13.05 \\
SMC center Dec. ($\degr$):          &   -72.85$\pm$0.10 & -72.83& -72.83 \\
SMC center distance (kpc):          &   59.0$\pm$1.5    & 62.44$\pm$0.47 & 62.44$\pm$0.47 \\ 
SMC center pmra (mas/yr):           &   0.75$\pm$0.10   & -0.743$\pm$0.027 &  -0.743$\pm$0.027\\
SMC center pmdec (mas/yr):          &   -1.26$\pm$0.05  & -1.233$\pm$0.012 & -1.233$\pm$0.012\\ 
SMC center systemic velocity (km s$^{-1}$):&   150.0$\pm$2.0   & 145.6$\pm$0.1 & 145.6$\pm$0.1 \\ 
SMC disk inclination ($\degr$):     &   70.0$\pm$10.0   & 81.9$\pm$0.7  & 58.4$\pm$1.4 \\
SMC disk position angle LON ($\degr$):&200.0$\pm$30.0   & 185.7$\pm$3.7  & 207.6$\pm$2.3 \\  
SMC disk rotation velocity (km s$^{-1}$):  &   25.0$\pm$5.0    & 10.0$\pm$5.0 & 10.0$\pm$5.0\\\hline

\end{tabular}
\end{table*}

\citet{hotaetal2024} recently analyzed Ultra Violet Imaging Telescope 
 \citep[UVIT;][]{tandonetal2017} data and {\it Gaia} Early DR3 data of stars in the shell region
in the north-eastern SMC. They split the star sample in three age groups: stars younger
than 150 Myr; stars with ages between 150 and 300 Myr, and stars older than 300 Myr. They
aimed to probe the kinematics of stars formed before and after the recent interaction
between the LMC and the SMC $\sim$ 250 Myr ago \citep{choietal2022}. They found no
apparent kinematic distinction between the three age groups, concluding that they did not
find any evidence of tidal perturbation or disruption in this part of the SMC. However, the
lack of distinction between proper motions of stars belonging to the north-eastern shell
with different ages is in agreement with the present findings. The stellar populations
formed in a region affected by galactic tides, as is the case of the outer north-eastern
region of the SMC, are kinematically perturbed by those tides. In the case of older
stellar populations, they were perturbed by tides because they were there when the tidal interactions 
occurred; and for the younger populations, the kinematics disturbance comes from the gas out of 
which they formed, which were sealed by that kinematic agitation too. As shown in Figure~\ref{fig5} 
(right panel) young clusters and old star populations have similar kinematics in the outer north-eastern
SMC shell ($\Delta$$V$ $<$ 60 km s$^{-1}$), while the motions of these clusters visibly differentiate 
($\Delta$$V$ $>$ 60 km s$^{-1}$) when they are compared to those of an orderly rotation disk 
extrapolated to the outer SMC regions.

I finally computed the anisotropy \citep{bennetetal2022}:\\

\begin{equation}
\beta = 1 - \frac{(\sigma V_r)^2 + (\sigma V_z)^2}{2 (\sigma V_\phi)^2},
\end{equation}

\noindent where $\sigma V_r$, $\sigma V_\phi$, and $\sigma V_z$ are the dispersion in the
three cylindrical velocity components, respectively. To obtained these quantities, I
first calculated the 3D velocity components of the five star clusters in cylindrical coordinates
with their respective uncertainties. Then, I used a maximum likelihood approach by 
optimizing the probability $\mathcal{L}$ that the five star clusters with velocities $V_i$ 
($V$ $\equiv$ $V_r$, $V_\phi$, $V_z$) and errors $e_i$ are  drawn from a population with mean $<V>$ and dispersion $\sigma$  
\citep[e.g.,][]{pm1993,walkeretal2006}, as follows:

\begin{equation}
\mathcal{L}\,=\,\prod_{i=1}^N\,\left( \, 2\pi\,(e_i^2 + \sigma^2 \, ) 
\right)^{-\frac{1}{2}}\,\exp \left(-\frac{(V_i \,- <V>)^2}{2(e_i^2 + \sigma^2)}
\right),
\end{equation}

\noindent where the errors on the mean and dispersion were computed from the respective 
covariance matrices. As a result, I obtained $\sigma V_r$, $\sigma V_\phi$, and $\sigma V_z$
with their corresponding uncertainties. Since the star clusters' velocities were
computed using as reference a young and an old SMC kinematic disk (see Table~\ref{tab2}),
I finally got two different $\beta$ values, namely: 0.06$\pm$0.01 (young disk) and
$-6.30$$\pm$0.10 (old disk), where the mean and dispersion come from averaging a thousand
values generated using Monte Carlo  for $\sigma V_r$= 12.2$\pm$0.1, $\sigma V_\phi$= 47.0$\pm$0.1, 
and $\sigma V_z$= 63.1$\pm$0.1 (young disk), and $\sigma V_r$= 13.9$\pm$0.5, $\sigma V_\phi$= 9.5$\pm$0.1, 
and $\sigma V_z$= 33.5$\pm$0.3 (old disk), respectively. I note that a negative value implies 
a low azimuthal dispersion, i.e. ordered rotation, which indicates similarity between the observations and 
the kinematic model. The resulting $\beta$ values reveal that the
kinematics of the five star clusters in the outer north-eastern is more similar to that
of the old SMC disk rather than to that of the young disk.

\section{Conclusions}

The strength of tidal effects caused by the interaction between galaxies 
should be reasonably measured by the level of kinematic disturbance of the
outermost galactic regions with respect to that in the innermost ones. This work
is aimed at confirming the above statement for the outer north-eastern region of the
SMC, where is known that a recent stellar structure formed because of the
interaction with the LMC \citep{martinezdelgadoetal2019,piatti2022d,dhanushetal2025}.
Since a number of young ($\la$ 200 Myr) star clusters also formed in this region during
the interaction, I used a sample of them as kinematic 
tracers of that region and computed their residual velocities ($\Delta$$V$).

To obtain $\Delta$$V$ values for HW~64, IC~1655, and IC~1660, three star clusters
formed in the north-eastern shell, I performed Gemini
GMOS-S multi-object observations, from which the mean star clusters' radial velocities
were derived for the first time. The mean clusters' proper motions were also
derived using {\it Gaia} DR3 data. The 3D space velocity components were calculated
employing the derived mean proper motions and radial velocities, star clusters'
heliocentric distance taken from the literature and the coordinates framework
defined in \citet{vdmareletal2002}. The modulus of the vector difference
between the cluster's velocity and that of a point in the SMC rotating disk at the cluster distance
from the SMC center represents $\Delta$$V$. The
resulting $\Delta$$V$ values, to which I added those of two star clusters belonging 
to the north-eastern shell with the necessary information available (HW~56 and NGC~458),
led me to conclude:\\

$\bullet$ When the SMC rotating disk fitted by \citet{piatti2021b} using star clusters
older than $\sim$ 1 Gyr is used as reference, the five studied star clusters would not
seem to be kinematically perturbed. Their $\Delta$$V$ values are mostly $<$ 60 km s$^{-1}$,
an upper limit found by \citet{piatti2021f} for the residual velocity of the SMC
 main body. The apparent lack of kinematic disturbance would seem to be the result of
employing as reference a rotating disk that by itself is perturbed. A similar behavior is
found if a rotating disk traced by old stellar populations is used.\\

$\bullet$ However, if an orderly rotating SMC disk fitted using young stellar populations
mainly distributed in the innermost galaxy regions is used as reference \citep[e.g.,][]{dhanushetal2025}, 
the star clusters exhibit a clear residual velocity gradient, in the
sense that the farther a cluster from the SMC center (in this case also implies closer to the
LMC), the larger its residual velocity  (see, Figure ~\ref{fig1append}). This outcome is a 
clear evidence that galactic tides
kinematically affects the outer galactic regions.\\

$\bullet$ The above results are in excellent agreement with the recent study by
\citet{dhanushetal2025}, who performed the computation of different SMC rotation disk parameters
employing a comprehensive sample of different stellar populations. Indeed, their results
point to the existence of as many different rotation disk as galaxy stellar component are used,
and highlight the prevailing orderly rotation motion in the innermost galaxy regions
with respect to the outer ones, which are affected by tides. \\

$\bullet$ The results found in this work confirm that LMC tides kinematically 
affect the outer SMC regions, independently of the age of the stellar populations that populate 
those regions. This implies that the epoch of close interaction between both Magellanic Clouds 
cannot be uncovered from the kinematics of the outer SMC regions.

\begin{acknowledgements}
 I thank the referee for the thorough reading of the manuscript and
timely suggestions to improve it.

Based on observations obtained at the international GEMINI Observatory, a program of NSF NOIRLab, which is managed by the Association of Universities for Research in Astronomy (AURA) under a cooperative agreement with the U.S. National Science Foundation on behalf of the GEMINI Observatory partnership: the U.S. National Science Foundation (United States), National Research Council (Canada), Agencia Nacional de Investigaci\'{o}n y Desarrollo (Chile), Ministerio de Ciencia, Tecnolog\'{i}a e Innovaci\'{o}n (Argentina), Minist\'{e}rio da Ci\^{e}ncia, Tecnologia, Inova\c{c}\~{o}es e Comunica\c{c}\~{o}es (Brazil), and Korea Astronomy and Space Science Institute (Republic of Korea).

This work has made use of data from the European Space Agency (ESA) mission
{\it Gaia} (\url{https://www.cosmos.esa.int/gaia}), processed by the {\it Gaia}
Data Processing and Analysis Consortium (DPAC,
\url{https://www.cosmos.esa.int/web/gaia/dpac/consortium}). Funding for the DPAC
has been provided by national institutions, in particular the institutions
participating in the {\it Gaia} Multilateral Agreement.

Data for reproducing the figures and analysis in this work will be available upon request
to the author.

\end{acknowledgements}

%\bibliographystyle{aa}
%\bibliography{paper} % if your bibtex file is called paper.bib

%\input{paper.bbl}

\begin{appendix}
\onecolumn

\section{Residual velocities}

\begin{figure}[h!]
\includegraphics[width=\textwidth/3]{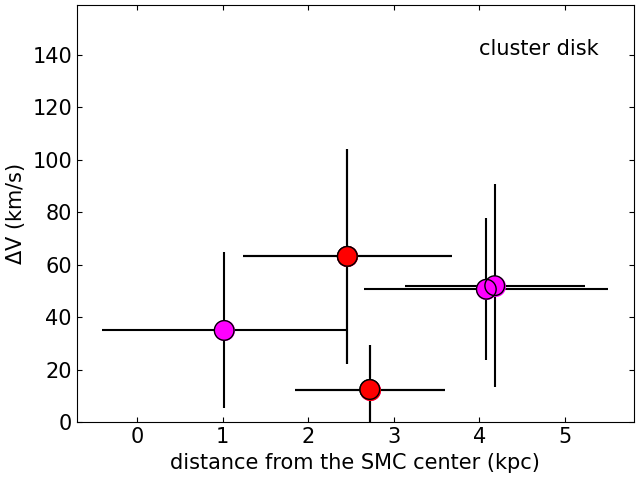}
\includegraphics[width=\textwidth/3]{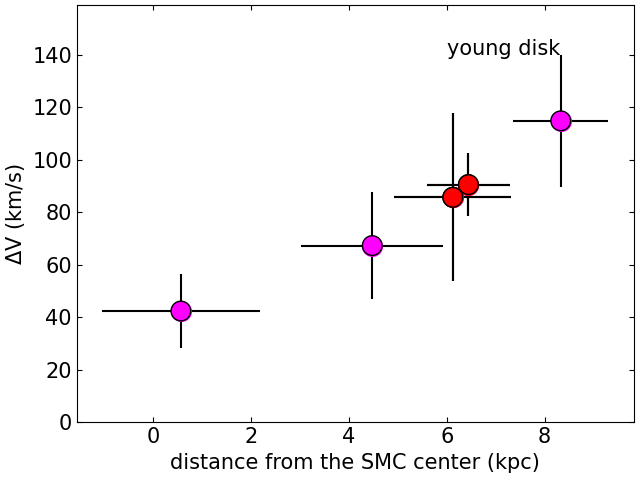}
\includegraphics[width=\textwidth/3]{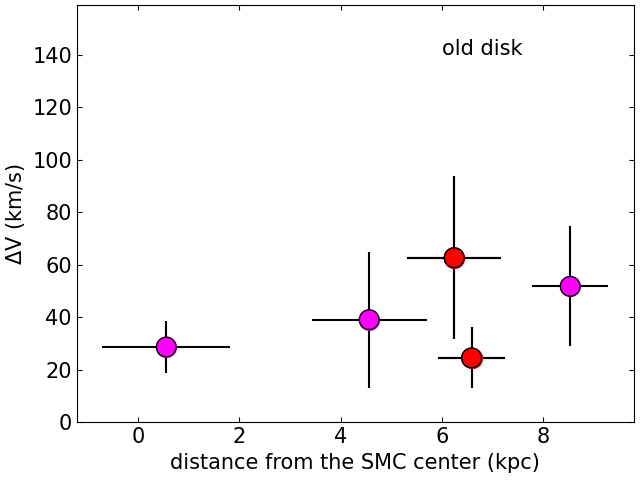}
\caption{Same as Figure~\ref{fig5} using 3D distances from the SMC center,
calculated using the models' parameters of Table~\ref{tab2}.}
\label{fig1append}
\end{figure}

\section{Spectroscopic data}

\begin{table}[!ht]
\caption{}
\label{tab1a}
%\small
\begin{tabular}{lccccrr}
\hline\hline
Star  & R.A. 	    & Dec.            & $g$ & $g-i$  & S/N & RV\hspace{0.8cm} \\
        & ($\degr$)  & ($\degr$)    & (mag) & (mag)  &       & (km s$^{-1}$)\hspace{0.5cm}  \\\hline
\multicolumn{7}{c}{HW~64}\\\hline
2&17.827&-71.322&17.41$\pm$0.01&1.05$\pm$0.06&  19.6 &137.05$\pm$8.91\\
4&17.766&-71.309&18.64$\pm$0.05&-0.47$\pm$0.05& 20.8 &154.98$\pm$6.69\\
5&17.809&-71.325&16.30$\pm$0.01&0.01$\pm$0.01&  27.6 &158.18$\pm$3.10\\
7&17.782&-71.322&15.49$\pm$0.04&-0.19$\pm$0.05& 51.5 &145.70$\pm$16.33\\
12&17.763&-71.332&17.43$\pm$0.04&-0.50$\pm$0.04&18.7 &138.33$\pm$19.95\\
13&17.740&-71.328&18.74$\pm$0.01&1.09$\pm$0.05& 13.0 &189.48$\pm$11.17\\
16&17.675&-71.316&15.61$\pm$0.01&-0.53$\pm$0.01&54.3 &156.21$\pm$1.58\\
17&17.753&-71.345&16.57$\pm$0.04&0.48$\pm$0.06& 27.7 &127.36$\pm$10.50\\
18&17.646&-71.314&17.99$\pm$0.02&0.76$\pm$0.03& 14.8 &133.24$\pm$11.11\\
19&17.784&-71.361&17.56$\pm$0.02&-0.48$\pm$0.02&14.2 &101.68$\pm$17.38\\
20&17.701&-71.338&19.39$\pm$0.02&0.92$\pm$0.02&  5.0 &165.53$\pm$7.32\\
21&17.687&-71.338&17.20$\pm$0.14&-0.60$\pm$0.20&24.3 &144.49$\pm$10.01\\
22&17.692&-71.341&19.81$\pm$0.12&-0.48$\pm$0.18&23.7 &190.41$\pm$4.53\\
23&17.682&-71.342&18.34$\pm$0.07&1.13$\pm$0.08& 13.2 &212.62$\pm$4.57\\
24&17.755&-71.368&18.72$\pm$0.01&0.81$\pm$0.02& 11.1 &145.13$\pm$10.71\\
25&17.630&-71.332&18.29$\pm$0.01&1.36$\pm$0.02& 14.6 &207.26$\pm$15.25\\
26&17.710&-71.362&17.42$\pm$0.02&-0.57$\pm$0.02&27.1 &164.77$\pm$3.41\\
27&17.731&-71.371&18.49$\pm$0.08&-0.44$\pm$0.08&10.7 &132.59$\pm$9.69\\
28&17.622&-71.338&16.33$\pm$0.04&-0.53$\pm$0.04&49.2 &160.44$\pm$7.77\\
30&17.637&-71.348&19.37$\pm$0.02&0.67$\pm$0.02&  6.8 &107.08$\pm$5.69\\
31&17.691&-71.367&17.94$\pm$0.02&-0.57$\pm$0.03&21.8 &44.09$\pm$14.84\\
32&17.740&-71.386&17.08$\pm$0.02&-0.31$\pm$0.03&28.8 &124.18$\pm$22.67\\
33&17.661&-71.364&17.26$\pm$0.03&1.12$\pm$0.04& 22.2 &177.35$\pm$8.66\\
34&17.676&-71.371&19.39$\pm$0.02&-0.47$\pm$0.02& 6.5 &154.39$\pm$3.29\\
35&17.635&-71.363&16.64$\pm$0.01&-0.56$\pm$0.02&64.2 &162.87$\pm$1.80\\
36&17.596&-71.353&18.68$\pm$0.03&-0.50$\pm$0.04&11.6 &89.73$\pm$4.91\\
39&17.609&-71.366&19.11$\pm$0.01&0.75$\pm$0.01&  8.9 &166.25$\pm$15.58\\
41&17.574&-71.361&19.95$\pm$0.01&-0.44$\pm$0.02&14.0 &162.79$\pm$22.36\\
42&17.654&-71.389&17.92$\pm$0.02&-0.54$\pm$0.04&15.8 &143.56$\pm$15.28\\\hline
\multicolumn{7}{c}{IC~1655}\\\hline
 1&17.841&-71.323&18.21$\pm$0.01&-0.54$\pm$0.01&25.9 &45.16$\pm$9.09\\
4&17.860&-71.328&18.12$\pm$0.01&-0.57$\pm$0.01& 20.2 &141.42$\pm$8.21\\
5&17.868&-71.336&17.51$\pm$0.02&-0.45$\pm$0.02& 26.3 &103.06$\pm$8.26\\
6&17.874&-71.360&17.69$\pm$0.02&-0.46$\pm$0.02& 16.6 &158.46$\pm$9.97\\
8&17.887&-71.347&17.31$\pm$0.01&1.08$\pm$0.03&  21.8 &166.57$\pm$4.04\\
9&17.894&-71.329&18.19$\pm$0.01&-0.57$\pm$0.01& 18.6 &144.72$\pm$13.26\\
10&17.900&-71.334&16.90$\pm$0.02&-0.48$\pm$0.03&36.3 &163.45$\pm$9.15\\\hline
\end{tabular}
\end{table}

\setcounter{table}{0}
\begin{table}[h!]
\caption{}
\label{tab1a}
%\small
\begin{tabular}{lccccrr}
\hline\hline
Star  & R.A. 	    & Dec.            & $g$ & $g-i$  & S/N & RV\hspace{0.8cm} \\
        & ($\degr$)  & ($\degr$)    & (mag) & (mag)  &       & (km s$^{-1}$)\hspace{0.5cm} \\\hline
\multicolumn{7}{c}{IC~1660}\\\hline
11&17.911&-71.343&18.60$\pm$0.01&-0.43$\pm$0.02&19.7 &142.34$\pm$3.53\\
13&17.922&-71.346&17.75$\pm$0.02&-0.54$\pm$0.02&25.1&136.07$\pm$10.60\\
15&17.937&-71.292&17.94$\pm$0.03&-0.52$\pm$0.03&18.2 &167.92$\pm$12.41\\
16&17.944&-71.361&17.87$\pm$0.01&-0.54$\pm$0.02&21.1 &120.98$\pm$21.75\\
17&17.949&-71.333&18.22$\pm$0.01&-0.51$\pm$0.01&21.0 &85.19$\pm$3.66\\
18&17.955&-71.324&18.94$\pm$0.01&0.73$\pm$0.01& 12.7 &166.60$\pm$2.10\\
19&17.962&-71.323&17.74$\pm$0.01&-0.59$\pm$0.01&28.8 &177.32$\pm$8.04\\
20&17.969&-71.318&17.21$\pm$0.01&1.09$\pm$0.22& 23.8 &179.81$\pm$1.28\\
21&17.975&-71.324&17.43$\pm$0.01&1.07$\pm$0.07& 20.1 &216.15$\pm$11.00\\
22&17.981&-71.331&17.87$\pm$0.01&-0.49$\pm$0.01&18.1 &72.33$\pm$6.26\\
23&17.991&-71.332&18.60$\pm$0.01&0.96$\pm$0.01& 16.6 &204.65$\pm$3.34\\
24&17.998&-71.320&18.31$\pm$0.03&-0.54$\pm$0.03&15.1 &204.12$\pm$4.56\\
25&18.006&-71.322&18.08$\pm$0.01&-0.54$\pm$0.01&20.4 &150.19$\pm$5.37\\
26&18.013&-71.328&17.19$\pm$0.02&0.86$\pm$0.05& 21.2 &202.44$\pm$3.78\\
28&18.026&-71.352&18.12$\pm$0.01&1.39$\pm$0.01& 13.3 &228.74$\pm$7.21\\
30&18.039&-71.344&16.94$\pm$0.01&1.01$\pm$0.01& 25.2 &153.60$\pm$2.37\\
34&18.069&-71.290&17.63$\pm$0.01&0.89$\pm$0.02& 19.2 &185.97$\pm$10.59\\
35&18.075&-71.315&18.80$\pm$0.01&1.14$\pm$0.01&  9.5 &212.77$\pm$4.66\\
36&18.080&-71.360&16.57$\pm$0.02&-0.45$\pm$0.03&40.8 &154.33$\pm$7.99\\
37&18.087&-71.318&19.28$\pm$0.05&-0.44$\pm$0.05& 8.1 &191.58$\pm$17.89\\
38&18.093&-71.339&19.04$\pm$0.01&1.05$\pm$0.01& 17.0 &157.08$\pm$3.80\\
40&18.106&-71.291&18.77$\pm$0.01&0.96$\pm$0.01&  9.7 &167.59$\pm$5.83\\
41&18.116&-71.354&18.02$\pm$0.01&-0.52$\pm$0.01&24.5 &148.43$\pm$11.85\\
1&18.104&-71.719&17.42$\pm$0.01&1.09$\pm$0.03&  17.6 &195.45$\pm$4.10\\
2&18.113&-71.719&18.40$\pm$0.01&1.13$\pm$0.01&   9.7 &191.78$\pm$9.69\\
3&18.058&-71.738&18.18$\pm$0.01&-0.52$\pm$0.01& 10.8 &129.19$\pm$1.86\\
4&18.089&-71.731&19.34$\pm$0.04&-0.41$\pm$0.11&  5.1 &77.76$\pm$6.92\\
5&18.114&-71.729&17.09$\pm$0.01&1.72$\pm$0.01&  17.0 &231.68$\pm$7.70\\
6&18.179&-71.712&18.78$\pm$0.01&1.30$\pm$0.01&  29.5 &141.02$\pm$14.88\\
7&18.064&-71.747&18.94$\pm$0.01&0.74$\pm$0.02&  25.4 &190.74$\pm$4.14\\
9&18.063&-71.752&17.63$\pm$0.01&1.92$\pm$0.01&  15.3 &298.81$\pm$20.94\\
12&18.199&-71.722&19.36$\pm$0.07&-0.41$\pm$0.07&16.7 &75.90$\pm$2.96\\
13&18.121&-71.747&17.76$\pm$0.01&0.86$\pm$0.01& 14.6 &148.85$\pm$1.37\\
18&18.182&-71.744&16.81$\pm$0.01&0.48$\pm$0.01& 30.5 &146.47$\pm$8.29\\
20&18.132&-71.763&18.03$\pm$0.01&1.43$\pm$0.04& 11.9 &122.08$\pm$3.80\\
21&18.160&-71.758&17.24$\pm$0.01&1.01$\pm$0.01& 22.4 &149.55$\pm$6.10\\
22&18.158&-71.761&15.63$\pm$0.07&0.53$\pm$0.18& 47.3 &165.72$\pm$4.77\\
23&18.168&-71.761&16.33$\pm$0.01&1.08$\pm$0.09& 25.4 &177.73$\pm$3.39\\
24&18.148&-71.770&17.03$\pm$0.01&0.68$\pm$0.04& 20.3 &92.80$\pm$11.74\\
25&18.211&-71.755&18.94$\pm$0.09&-0.38$\pm$0.09&21.4 &-29.76$\pm$5.50\\
30&18.189&-71.774&16.38$\pm$0.01&0.16$\pm$0.07& 43.5 &107.13$\pm$10.74\\
38&18.132&-71.818&17.71$\pm$0.01&1.58$\pm$0.01& 23.5 &154.97$\pm$6.76\\
40&18.285&-71.780&19.34$\pm$0.13&-0.47$\pm$0.21&18.7 &88.01$\pm$23.89\\\hline

\end{tabular}
\end{table}

\twocolumn
\end{appendix}

\end{document}